\documentclass[prd,aps,floats,twocolumn,preprintnumbers]{revtex4}

\usepackage{amsmath}
\usepackage{epsfig}
\usepackage{amssymb}

\newcommand{\be}{\begin{equation}}
\newcommand{\ee}{\end{equation}}
\newcommand{\bea}{\begin{eqnarray}}
\newcommand{\eea}{\end{eqnarray}}
\newcommand{\lag}{{\cal L}}

\newcommand{\mpl}{M_{\rm P}}

\def\vf{\varphi}
\def\p{\partial}

\def\om{\omega}
\def\N{\nabla}

\begin{document}

\setlength{\unitlength}{1mm}
 
\title{Baryogenesis after Hyperextended Inflation}

\author{Antonio De Felice and Mark Trodden}

\affiliation{Department of Physics, Syracuse University, Syracuse, NY
13244-1130, USA.\\ 
{\tt defelice@phy.syr.edu, trodden@phy.syr.edu}}

\begin{abstract}
We study a baryogenesis mechanism operating in the context of hyperextended inflation and making use of a coupling between the scalar field and a standard model global current, such as $B$ or $B-L$. The
method is efficient at temperatures at which these currents are not conserved due to some higher dimensional operator. The particle physics and cosmological phenomenology are discussed. We consider constraints stemming from nucleosynthesis and solar system experiments.
\end{abstract}

\maketitle

\section{Introduction}

In modern cosmology, scalar fields play a key role in our picture of the model of the universe on large scales. In particular rolling scalar fields minimally coupled with gravity are important in cosmology since they have been used, for example, to realize inflation and quintessence~\cite{Guth:1980zm,
Linde:1981mu, Albrecht:1982wi, Caldwell:1997ii, Zlatev:1998tr,Turner:1998ex, Weiss:1987xa, Ratra:1987rm, Wetterich:fm, Peebles:1987ek,Frieman:1995pm, Coble:1996te,Armendariz-Picon:2000ah}.

Recently a number of minimal models have been proposed, in which a single scalar field is used to achieve several different goals. One example is provided by models that explain inflation and quintessence by means of the same scalar field~\cite{Peebles:1998qn}. One can also try to unify other sectors of the theory, for example by realizing baryogenesis, or by providing a description of dark matter via a scalar field by an opportune choice of its equation of state~\cite{Bento:2002ps, Bilic:2001cg}. An alternative way to unify dark energy and dark matter in a single model would be to introduce a scalar field that behaves differently according to the local density of matter~\cite{Khoury:2003rn}.

For minimally coupled fields, the dark energy behavior is described by a suitable choice for the potential of the scalar field. We may generalize this picture to scalar fields which are non-minimally coupled with the gravity sector. Such models arise naturally, for example, in the context of extra-dimensional models, for which the radion field describing the size of the extra dimensions is a scalar-tensor (ST) scalar field. These ST fields have been exploited to model dark energy behavior in a quite large variety of models~\cite{Faraoni:2001tq, Gunzig:2000kk, Riazuelo:2001mg, Uzan:1999ch, Perrotta:2002bk, Perrotta:2002sw, Perrotta:1999am, Chiba:2001xx, Chen:1999qh, Torres:2002pe, Santos:1996jc,Carroll:2004hc,Perivolaropoulos:2003we}  and are a natural complement to other modified gravity theories which lead to cosmic 
acceleration ~\cite{Carroll:2003wy, Nojiri:2003ft, Chiba:2003ir,Flanagan:2003iw,Carroll:2004de}. 

If the ST field is a radion, then the potential that stabilizes the extra dimensions causes the field to become sufficiently massive that the solar system constraints do not apply. However, for general ST theories, in the absence of a strong potential, the main constraints on light scalar 
fields come from solar system experiments. 

In this paper we seek to connect the ST field to baryogenesis. Related work has used a minimally coupled scalar field to achieve inflation, baryogenesis and quintessence~\cite{DeFelice:2002ir}, or has coupled standard model currents to a derivative of the Ricci scalar to describe baryogenesis~\cite{Davoudiasl:2004gf}.

Here we are interested in the role that a scalar field in ST theories may play in the generation of the baryon asymmetry of the universe. The spectacular success of primordial nucleosynthesis requires that
there exists an asymmetry between baryons and anti-baryons in the universe at temperatures lower than 1 MeV. This is quantified by
\be
4\times10^{-10}\leq\eta\equiv\frac{n_B}s\leq7\times10^{-10}\ ,
\ee
where $n_B$ is the difference between the number density of baryons and that of anti-baryons and $s$ in the entropy density.

The outline of this paper is as follows. In Sec.~II we will review some of the details of scalar-tensor theories, derive the equations of motion and introduce the current bounds on these kinds of models. In Sec.~III we describe the ``hyperextended inflation'' model and how it arises in such theories. In Sec.~IV the behavior of the scalar-tensor field during the radiation era is discussed. In Sec.~V we will describe how scalar tensor theories may naturally yield a baryon asymmetry, by introducing a coupling between a ST scalar field and a global current of standard model, such as $B$ or $B-L$. 
The evolution of the ST scalar field during the matter dominated era is discussed in Sec.~VI. In
Sec.~VII we consider constraints on our model, coming from the coupling we introduced to generate the baryon asymmetry and we offer our comments and conclusions in Sec.~VIII.

\section{The Scalar-Tensor Action}

To begin with let us discuss a pure Brans-Dicke theory~\cite{Brans:sx,
Dicke:1961gz}. The Lagrangian density is 
\be
 \lag_{BD}=\sqrt{-g} \left[\vf\,R - \frac\omega\vf\,(\partial_\mu \vf)^2+ \lag_m\right] \ ,
\ee 
where $\vf$ is a scalar field with units of mass$^2$ and 
$\om$ is a constant. For $\om>3/2$ it is easy to
find a conformal transformation for the metric and a field
redefinition for the scalar field such that in the new frame, called the
Einstein frame, the Lagrangian becomes that of a scalar field minimally
coupled to gravity but non-minimally coupled to matter. 

Attempts to implement inflation within this theory are known as {\it extended inflation}.
However, such attempts are generally unsuccessful~\cite{Mathiazhagan:vi, La:za}, 
since if $\om$ is too small one obtains insufficient inflation and if $\om$ is too big then
unacceptably large fluctuations in the temperature of the Cosmic Microwave Background 
Radiation (CMB) are generated~\cite{Weinberg:1989mp, La:pn}. 

Since gravity in the solar system is well-approximated by General Relativity today, we
require that
\begin{equation}
\vf|_{\rm today} \equiv \vf_0 = \mpl^2\ .
\end{equation}
Further, the most current solar system experiments, performed by 
the Cassini-spacecraft~\cite{Cassini}, force $\om$ to be greater than
40000 (see e.g.~\cite{Will:2001mx, Reasenberg:ey}). 
Thus, for models of massless ST fields for which $\om$ is a
constant or approximately a constant, this constraint holds, 
even though the value of $\om$ may change on cosmological
scales~\cite{Clifton:2004st}. 

Comparing with General Relativity, the field $\vf$ plays the role of an effective 
dynamical Newton's constant and, as such, its variation is bounded 
by~\cite{Williams:1995nq,Dickey}
\begin{equation}
\left|\frac{\dot\vf}\vf\right|
\propto\left|\frac{\dot G_{\rm eff}}{G_{\rm eff}}\right|
<6\times10^{-12}{\rm\ yr}^{-1}\ .
\label{solar}
\end{equation}

\section{Models of Scalar-Tensor Inflation}

We wish to explore whether, within these scalar-tensor theories, the same field $\vf$ may be responsible for inflation and then, later, responsible for baryogenesis. We shall first review the {\it hyperextended inflation} (HI) model, which provides a viable scalar-tensor model of inflation.

As we made clear earlier, it is necessary to generalize the pure BD Lagrangian to achieve successful inflation. One way to do this is to promote $\om$ to be a general function of $\vf$~\cite{Barrow:1990nv, Steinhardt:1990zx}. The Lagrangian is then
\begin{equation}
\lag_{ST}=\vf\,R - \frac{\omega(\vf)}\vf\,(\partial_\mu \vf)^2+
16\pi\,\lag_m \ .
\label{azione}
\end{equation}
If $\dot\vf\to0$ then $\vf$ goes to a constant and we get a standard GR. We have not introduced any potential for the scalar field, although it might prove useful to introduce it at a later time in order to describe dark energy.

This model was introduced by Barrow and Maeda~\cite{Barrow:1990nv} and was analyzed in detail in~\cite{Liddle:1991am, Green:1996hk}. As we shall see later on, this model encounters problems because $\vf$ is required to roll a little too quickly. Nevertheless, it serves as an explicit example of how our model works in the context of ST theories. More realistic models of ST inflation are discussed in~\cite{Steinhardt:1990zx,Crittenden:cf}.

The action~(\ref{azione}) yields the equations of motion for the gravitational sector
\bea
\vf\,R_{\mu\nu} &-& \frac12\,g_{\mu\nu}\,\vf\,R +
\frac\om{2\vf}\,g_{\mu\nu}\,(\p\vf)^2-
\frac\om\vf\,\p_\mu\vf\,\p_\nu\vf \nonumber \\
&-& \N_\mu\N_\nu\vf+g_{\mu\nu}\,\Box\vf
=8\pi\,T_{\mu\nu}\ ,
\label{prima}
\eea
and
\be
R-\frac\om{\vf^2}\,(\p\vf)^2+\frac{\om'}\vf\,(\p\vf)^2
+2\,\frac\om\vf\,\Box\vf=0\ ,
\label{seconda}
\ee
where $T_{\mu\nu}$ is the energy-momentum tensor for matter, and a prime denotes a derivative with respect to $\vf$.

Combining the trace of~(\ref{prima}) with~(\ref{seconda}) then gives
\be 
\Box\vf=\frac{8\pi}{2\om+3}\,T-\frac{\om'}{2\om+3}\,(\p\vf)^2\ ,
\ee

In order to study the cosmological consequences, we adopt the spatially flat FRW ansatz for the metric
\begin{equation}
ds^2=-dt^2+a(t)^2\,[dr^2+r^2\,(d\theta^2+\sin^2\theta\,d\phi^2)]
\end{equation}
and represent matter by a perfect fluid with an equation of state $p=w\,\rho$, with constant equation of state parameter $w$. Assuming that $\vf$ is homogeneous, so that $\vf=\vf(t)$, leads to the following modified Friedmann equation
\begin{equation}
\left(\frac{\dot a}a\right)^{\!2}=\frac{8\pi}{3\vf}\,\rho-
\frac{\dot\vf}\vf\,\frac{\dot a}a+
\frac\omega6\left(\frac{\dot\vf}\vf\right)^{\!2}\ ,
\end{equation}
where $\vf$ satisfies
\begin{equation}
\ddot\vf+3\,\frac{\dot a}a\,\dot\vf=\frac1{3+2\,\om}\,
\bigl[8\pi\, \rho(1-3w)-\om'\,\dot\vf^2\bigr]
\label{ST_eq}
\end{equation}
and
\begin{equation}
\rho(a)=\rho_0 \left(\frac{a_0}{a}\right)^{3(1+w)} \ .
\end{equation}

Hyperextended inflation involves the above gravitational sector, plus an inflaton field $\Sigma$. The potential for $\Sigma$ is chosen to have a false vacuum in which the inflaton becomes trapped, drives inflation, then ultimately tunnels and rolls down to the true minimum.  During this process, true-vacuum bubbles nucleate and finally merge. In this sense, HI shares many of the features of Guth's original old inflation model~\cite{Guth:1980zm}.

At the onset of inflation, $\vf$ starts at very small values ($\vf\approx0$), and during inflation increases to values approximately equal to the value it has today, that is $\vf_x\approx\vf_0$, where a subscript $x$ denotes the value of quantities at the end of inflation and a subscript 0 denotes their
value today.  Furthermore, the function $\om(\vf)$ starts at small values and reaches, at the end of inflation, the value we have today, constrained to be $\om_0\geq40000$. In more detail, one assumes the form 
\begin{equation}
\om(\vf)=\om_i+\om_m\left(\frac\vf{\vf_0}\right)^{\!m}\ ,
\end{equation}
where $m$ is a positive integer and $\om_i$ and $\om_m$ are constants. In order for the theory to be consistent with the spectrum of fluctuations in the CMB and for inflation to last for a sufficient number of e-foldings, one requires
\begin{eqnarray}
\om_i&\sim& O(1)\ ,\\
\om_m&\geq&40000\ , \\
m&>&5\ .
\end{eqnarray}

As was pointed out by Liddle and Wands~\cite{Liddle:1991am}, this is the weak part of the model, because the transition from small to large values of $\vf$ must occur extremely rapidly and is difficult to explain in a natural way. In particular it is not known what kind of symmetry would keep all terms with $m\leq5$ small during the evolution of the field $\vf$.

As an initial condition we may set $\dot\vf$ equal to 0. At the end of inflation it is sufficient, for our purposes, to derive an approximate solution for $\vf$ by neglecting terms in $\ddot\vf$ and $\dot\vf^2$ in equation~(\ref{ST_eq}). This yields, for times $t\leq t_x$,
\begin{equation}
\vf(t) \approx
\vf_x\left[1+(2m+1)\,\frac{H_x}{\om_x}\,(t-t_x)\right]^{\!2/(2m+1)} \ .
\end{equation}
From this relation we can easily calculate both $\dot\vf$ and $\ddot\vf$ and check that the solution is consistent with the approximations we made to solve the equation. Furthermore, we find that, at the end of inflation,
\begin{equation}
\frac1{H_x}\,\frac{\dot\vf_x}{\vf_x}\approx\frac 2{\om_m}\ll1\ ,
\end{equation}
as required by~(\ref{solar}). The value of the temperature $T_x$ at the end of inflation can then be estimated by assuming instantaneous reheating, so that almost all the initial energy of the inflaton is converted into particles; that is 
\be
\rho_\Sigma (T_x) \approx T_x^4\ ,
\ee
For hyperextended inflation models this typically yields $T_x\sim10^{15}$ GeV, (see e.g.~\cite{Liddle:1991am}).

\section{The Radiation-Dominated Era}
Having described the main features of HI models, we now turn to the evolution of the relevant fields after inflation ends.

After inflation the universe enters a period of radiation domination, since ultra relativistic particles have been created through reheating and the collisions of true-vacuum bubbles. It is in this era, before nucleosynthesis, that our model of baryogenesis operates.

Since $\dot\vf$ is quite small at the end of inflation, the value of $\vf$ does not change very much, It is therefore a good approximation to take
\begin{equation}
\vf_x\approx\vf_0=\mpl^2\ .
\end{equation}
At this level of approximation we may keep only the largest term in the Friedmann equation, which becomes
\begin{equation}
H^2\approx\frac{8\pi}{3\,\mpl^2}\,\rho\ ,
\end{equation}
so that, in the first approximation, we recover GR with radiation source obeying $ \rho \approx T^4$ and $n \approx T^3$.

To find the equilibrium temperature, $T_{\rm th}$, note that, since particles are ultra-relativistic, we may approximate their cross section by $\sigma\sim \alpha^2 / T^2$, where $\alpha$ is a dimensionless coupling constant of order 0.1 or 0.01. Because $H\sim n\,\sigma$ in equilibrium, we therefore obtain
\begin{equation}
T_{\rm th}=\alpha^2\,\sqrt{\frac3{8\pi}}\,\mpl\sim10^{15}\ {\rm GeV}\ .
\end{equation}
This is the equilibrium temperature. Thus, as soon as inflation ends, the radiation is in equilibrium and baryogenesis can take place.

It is important for our analysis to know the behavior of the ST field $\vf$ during this era. Assuming all particles are ultra-relativistic, then the trace of the stress-energy tensor vanishes, and in equation~(\ref{ST_eq}) $\vf$ decouples from matter. We may then integrate equation~(\ref{ST_eq}) exactly, to obtain
\begin{equation}
\dot\vf=\dot\vf_x\left(\frac{a_x}a\right)^{\!3}
\left[\frac{3+2\,\om(\vf_x)}{3+2\,\om(\vf)}\right]^{\!1/2}.
\label{soluzio_impo}
\end{equation}
A Taylor expansion of the square root in~(\ref{soluzio_impo}) then leads to
\begin{equation}
\dot\vf
\approx
\dot\vf_x\left(\frac T{T_x}\right)^{\!3}\ .
\label{equaz_per_phi_dot}
\end{equation}
where we have used $a\propto T^{-1}$.

\section{Scalar-Tensor Baryogenesis}
We now turn to our
model. We are interested in the idea that $\vf$ may play a role in generating
the baryon asymmetry, and therefore we need to know how $\vf$ couples with
other fields.  We choose to couple it with standard model matter fields by a
dimensionless coupling constant and a mass scale that we will constrain later
by observations.

We consider terms in the effective Lagrangian of the form
\begin{equation}
\lag_{\rm eff}=\frac\lambda{M^2}\,\partial_\mu\vf\, J^\mu\ ,
\label{effe}
\end{equation}
where $\lambda$ is a coupling constant, $M<\mpl$ is the scale of the cutoff in
the effective theory and $J^{\mu}$ is the current corresponding to some
continuous global symmetry such as $B$ or $B-L$. We assume that $\vf$ is
homogeneous so that
\begin{equation}
\lag_{\rm eff} = \frac\lambda{M^2}\,\dot\vf\,\Delta n\equiv 
\mu(t)\,\Delta n\ .
\end{equation}
where $n=J^0$ is the number density corresponding to the global symmetry and
we can interpret $\lambda\dot\vf/M^2$ as an effective chemical potential
$\mu$.

The effective Lagrangian~(\ref{effe}) breaks CP and, furthermore, the standard
model, due to its chiral structure, is C-violating. In addition, the fact that
${<}\dot\vf{>}\neq0$ breaks $T$ symmetry, so that $CPT$ is broken. The model
of baryogenesis relevant here is {\it spontaneous baryogenesis}. This model
works in thermal equilibrium and was introduced for a scalar field, the
thermion, by Cohen and Kaplan~\cite{Cohen:1988kt}, who suggested that
spontaneous baryogenesis might take place during reheating. Recently
it was proposed~\cite{DeFelice:2002ir} that the quintessence scalar could
drive baryogenesis in a model termed {\it quintessential
baryogenesis}.

In equilibrium ($T<T_{\rm th}$), we calculate $\Delta n$ by using the
distribution functions at equilibrium for bosons or fermions as follows
\bea
\Delta n &=& \int \frac{g\,d^3p}{(2\pi)^3}
\,[f(E,\mu)-f(E,-\mu)] \nonumber \\
&\approx& \frac{g\,T^3}6\,\frac\mu T + 
O\!\left(\frac\mu T\right)=
\frac{g\,T^2}6\,\mu\ ,
\eea
where $g$ represents the number of degrees of freedom of the matter fields
coupled to $\vf$ and we have expanded $\Delta n$ to first order in
$\mu/T<1$. Using the value of $\mu$ we found earlier, we obtain
\begin{equation}
\Delta n(T;\mu) \approx \frac{\lambda\, g}{6\,M^2}\,T^2\,\dot\vf\ .
\label{delta}
\end{equation}

In order to produce a baryon asymmetry we require some, as yet unspecified, 
$B-L$ number violating processes. 
Below some temperature, $T_F$, these reactions
will become so rare that they will freeze out, and at that point baryon production
will cease and the baryon number, if the global current is $B-L$, will change no more. 
The final baryon asymmetry will then be given by
\begin{equation}
\eta_F \equiv \eta(T_F) \equiv \frac{\Delta n}s(T_F) = 
 \frac{15}{4\pi}\,\frac g{g_*}\, 
\frac{\lambda\,\dot\vf_F}{T_F\,M^2}\ ,
\label{eta_finale}
\end{equation}
where we have used that the entropy density is $s = 2\pi g_*\,T^3 /45$ .
Baryogenesis is therefore effective at temperatures $T_{\rm th} > T>T_F$,
with corresponding scale factors $a_{\rm th}<a<a_F$.

In order to calculate $\eta_F$ we need to know the dynamics of $\vf$ during
the radiation-dominated era. We examined this in the previous section but, to make
sure that analysis holds, we must investigate the
possible back-reaction effects of the effective Lagrangian we introduced in
equation~(\ref{effe}) on $\vf$. The equation of motion of $\vf$ becomes
\bea
\ddot\vf+3H\,\dot\vf&=&
\frac{8\pi}{2\om+3}\,\rho(1-3w)-\frac{\om'}{2\om+3}\,\dot\vf^2\nonumber\\
&-& \frac{\lambda}{M^2}\frac2{2\om+3}\,\Delta n\,\dot\vf\nonumber\\
&-& \frac{\lambda}{M^2}\frac\vf{2\om+3}
\left[\frac{d(\Delta n)}{dt}+3\,H\,\Delta n\right] .
\eea
Using~(\ref{delta}) we find
\bea
\ddot\vf\,(1+\Theta) &+& 3\,H\,\dot\vf\,\left(1+\frac{\Theta}{3}\right) \nonumber \\
&+& \frac{\om'\,\dot\vf^2}{2\om+3}
\left[1+\frac{2(2\om+3)\,\Theta}{\vf\,\om'}\right] \nonumber \\
&=& \frac{8\pi\,(\rho-3p)}{2\om+3}\ ,
\eea
where
\be
\Theta = \frac g{6(2\om+3)}\left(\frac{\lambda\,\vf}{M^2}\right)^{\!2}
\frac{T^2}\vf\ .
\ee

Now, as an example let us take $\lambda\,\vf/M^2\sim200$. As we shall see in section~VII, this is as large as it is allowed by current constraints. Also $T^2/\vf\sim (T/\mpl)^2\lesssim10^{-8}$. Further, $(3+2\,\om)\sim80000$ and $\vf\,\om'\approx m\,\om_m\sim40000\,m$. It is therefore clear that $\Theta\ll1$ and we may safely neglect the back-reaction on $\vf$ from the coupling we have introduced.

\subsection{Baryon Number Violation}
Since baryon number is a global symmetry of the standard model, operators that
violate baryon number should appear as non renormalizable operators. In order
to calculate $\eta_F$ in equation~(\ref{eta_finale}) we need to find $T_F$,
the freeze out temperature for the baryon number violating interactions. To do
this requires an explicit form for the effective Lagrangian that violates $B$
or $B-L$. One simple example is to introduce an effective 4-fermion
operator
\be
\lag_{B-L}=\frac{g'}{M_X^2}\,\bar\psi_1\,\psi_2\,\bar\psi_3\,\psi_4\ ,
\label{opera}
\ee
where each $\psi_i$ represents a standard model Fermi field. $M_X$ is a mass
scale which may be thought of as the mass of the gauge boson mediating the
force responsible for $B-L$ violating processes. The operator~(\ref{opera})
would be obtained by integrating out such a boson to obtain the low-energy
effective theory.

The rate for the $B-L$ violating processes is given by $\Gamma_{B-L} =
{<}n\,\sigma v{>}$, where $\sigma$ is the total cross section, $n\sim T^3$ is
the number density, $v\sim1$ is the relative velocity; and $<>$ represents a
thermal average. For $T<M_X$ we may write $\sigma\sim g'^2\,T^2/M_X^4$ giving
\be
\Gamma_{B-L} \sim \frac{g'^2}{M_X^4}\,T^5\ .
\ee
If the universe expands at a rate faster than this, then the relevant particle
interactions are out of equilibrium. This happens above a temperature $T_F$
given by
\be
\Gamma_{B-L}(T_F)=H(T_F)\ .
\label{rata}
\ee
Equation~(\ref{rata}) implies that
\begin{equation}
T_F\sim\left[\sqrt{8\pi/3}\,\tilde g^{-2}\right]^{1/3}
\left(\frac{M_X}{\mpl}\right)^{\!1/3} M_X\ .
\label{freeze}
\end{equation}

Combining equation~(\ref{equaz_per_phi_dot}) with~(\ref{eta_finale}),
we then find
\begin{equation}
\eta_F = \sqrt{\frac{73}{2\pi}}\,
\frac g{g_*}\, \frac{T_x}{\mpl}
\left(\lambda\,\frac{\vf_x}{M^2}\right)
\left(\frac1{H_x}\,\frac{\dot\vf_x}{\vf_x}\right)
\left(\frac{T_F}{T_x}\right)^{\!2} .
\label{baryo}
\end{equation}
Typically, for the models in this paper, we have $T_x\sim10^{15}$ GeV, so that $T_x/\mpl\sim 10^{-4}$. Further, the ratio $g/g_*$ is of order $10^{-2}$, and $(73/2\pi)^{1/2}$ is approximately 3.

Note that $\vf_x \sim \mpl^2$ and, as we shall see later on when we discuss constraints on our model,
$\lambda\,\vf_x/M^2$ can be quite large, but is constrained to be less than 200
if variations in Newton's constant are close to current bounds. Also, 
although $H_x^{-1}\,\dot\vf_x/\vf_x$ is
model dependent, we saw earlier that, for hyperextended inflation, this
quantity is of order $2/\om_m$, where $\om_m\sim40000$.

Finally, from equation~(\ref{baryo}), since our goal is to obtain $\eta_F$ of
order $10^{-10}$, we require $T_F/T_x\lesssim5\times10^{-2}$ or
$T_F\sim10^{13}$ GeV. By using equation~(\ref{freeze}), again assuming that
$T_x\sim10^{15}$ GeV, we find that a mass $M_X\sim10^{14}, 10^{15}$ GeV is
required to have an $\eta_F$ which can fulfill nucleosynthesis
requirements. This value is very close to the mass scale that comes naturally
in contexts like Grand Unification Theories and Supersymmetry, so that the
operator~(\ref{opera}) that violates $B-L$ at GUT scale, just tells us that
some new physics appears at that scale. In this sense the choice of a
$B$ or $B-L$-violating operator seems to be a natural one.

It should be pointed out that, since $T_F\gg T^{EW}\sim100$ GeV, anomalous
electroweak processes at finite temperature may affect this result. However if
we generate a $B-L$ asymmetry the amount produced by ST baryogenesis will not
be altered.

\section{Evolution During the Matter-Dominated Era}
\begin{figure}[ht]
\centering
\includegraphics[width=2.5in]{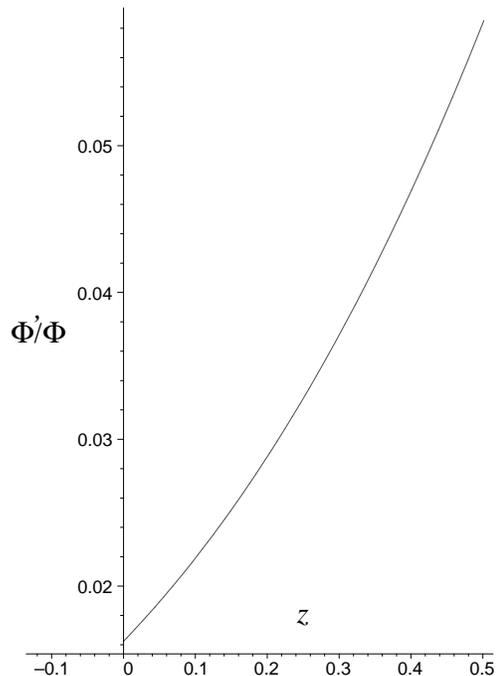}
\caption{$\Phi^{-1}\,d\Phi/d\tau$ as a function of the redshift $z$.}
\label{diffa}
\end{figure}

In order to put constraints on our model it is useful to know how $\vf$ evolves
during matter domination. Consider again~(\ref{ST_eq})
\begin{equation}
\ddot\vf+3\,\frac{\dot a}a\,\dot\vf=\frac1{3+2\,\om}
\left[8\pi\,(\rho-3p)-\frac{d\om}{d\vf}\,\dot\vf^2\right]\ .
\label{mat_eq}
\end{equation}
Let us change variables, $\tau=H_0\,(t-t_0)$, $\vf=\vf_0\,\Phi$, $\tilde a = a/a_0$
and rewrite this as
\be
\Phi''+3\,\frac{\tilde a'}{\tilde a}\,\Phi'=
\frac3{3+2\,\om}\,\tilde a^{-3}
-\frac1{3+2\,\om}\frac{d\om}{d\Phi}\,\Phi'^2\ ,
\label{num_vf}
\ee
where we have neglected any kind of matter but dust, and have denoted by a prime differentiation with
respect to $\tau$ (for a detailed analysis of ST theories describing dark
energy with matter included see~\cite{Catena:2004ba,Catena:2004pz}). The Friedmann
equation becomes
\be
\frac{\tilde a'}{\tilde a}=-\frac12\,\frac{\Phi'}\Phi+
\left[\frac{2\om+3}{12}\left(\frac{\Phi'}\Phi\right)^{\!2}+
\frac1{\Phi\,\tilde a^3}\right]^{\!1/2}\ .
\label{num_fr}
\ee
We solve~(\ref{num_vf}) and~(\ref{num_fr}) subject to $\tilde a(0)=1$, $\Phi(0)=1$ and $\Phi'(0)=H_0^{-1}\,\dot\vf_0/\vf_0$ ($\tau=0$ corresponds to today with $\Phi'(0)\approx0.057$.
A particularly useful result is shown in figure~(\ref{diffa}), where $\Phi'/\Phi$ is plotted against $z$,
showing that the derivative of $\vf$ keeps decreasing during matter
domination.

\section{Constraints}\label{constr}
After the radiation and matter eras, $\vf$ keeps evolving with a small $\dot\vf$. As a result, $\om(\phi)$ is approximately constant and $\vf$ becomes more and more similar to a pure BD scalar field. Therefore, if the field is light, we have a light scalar field with a small coupling to standard model fields. Scalar fields of this kind have potentially observable consequences. The coupling~(\ref{effe}) we took
as the basis for our model can be written as
\be
\lag=-\lambda\,\frac\vf{M^2}\,\partial_\mu\,J^\mu\ .
\label{effe2}
\ee

for us, $J^{\mu}$ corresponds, for example, to the $B-L$ symmetry. However, if we allow
such a coupling there is no reason not to allow a similar coupling to the electromagnetic
current. This kind of coupling has been studied by Carroll~\cite{Carroll:1998zi} in
detail in the context of quintessence scalar fields.

An analogous constraint will apply also in our model and can be found by following the 
similar analysis in~\cite{DeFelice:2002ir}. It is straightforward to show that the bound becomes
\be
\lambda\,\frac{n_f\,g'^2}{32\pi^2} \leq 3\times10^{-2}
\,\frac{M^2}{|\Delta\vf|}\ ,
\label{contro}
\ee
where $|\Delta\vf|$ is the variation of the field during the last half a
redshift of the universe ($z\leq1/2$) and can be bounded using experimental constraints 
on the time variation of Newton's constant (see equation~(\ref{solar})).

Setting 
\be
|\Delta\vf|=\vf_0 \left(\frac{1}{H_0}\right)\left(\frac{\dot\vf_0}{\vf_0}\right)\, (H_0\, |\Delta
t|)+O(\Delta t^2) \ ,
\ee 
with $z\sim H_0\,|\Delta t|$, and using $H_0=100h$ km/s/Mpc, 
with $h=0.72\pm0.08$, we find
\be
|\Delta\vf|\sim \mpl^2\,\frac1{H_0}\,\frac{|\dot\vf_0|}{\vf_0}\,z\ .
\ee
Using this value, the
constraint~(\ref{contro}) for our model can be approximately written as
\begin{equation}
\lambda\left(\frac{|\Delta\vf|}M^2\right)\leq 8 \ .
\end{equation}
If variations in Newton's constant are close to current limits, then this can also be translated 
into a bound $\lambda\,\vf_x/M^2 \lesssim 200$. Of course, if Newton's constant varies less 
dramatically, this constraint is considerably weakened. 

There may be other constraints on specific implementations of this model, such as those 
arising from proton-decay. However, it is worth pointing out that it is also possible that the operator~(\ref{opera}) is responsible only for the production of a leptonic asymmetry (see,
for example,~\cite{Zhang:1993vh}). 
Subsequently, at electroweak scale, sphaleron transitions will then reprocess lepton number 
and convert a fraction of it into baryon number.

\section{Conclusions}
Attempts to go beyond General Relativity frequently lead to theories containing 
fields which are non-minimally coupled to gravity. As a simple, minimal extension 
of gravity, we may consider theories of a single scalar field with a
generalized coupling to gravity, known as scalar-tensor theories. Such
theories arise in a natural way in string theory and more general extra-dimensional models,
and can be used to describe inflation, hyperextended inflation and dark
energy.

In this paper we have investigated the possibility that such a scalar field
may be important in other sectors of cosmology, in particular in explaining the origin
of the baryon asymmetry of the universe. We have
considered a non-minimally coupled scalar field, coupled to standard model
fields and demonstrated that this coupling may ultimately lead to a non-zero
baryon asymmetry in a quite general way.

The model provides a natural link between inflation and baryogenesis and appears
to be less constrained than related ideas~\cite{DeFelice:2002ir} involving minimally-coupled fields.
We have left a number of questions
unanswered, for example the stability of this model under quantum corrections, and may 
return to them in a future work. 

\acknowledgments 
ADF would like to thank the nice environment of TASI 2004, and the
University of Colorado at Boulder for
hospitality. The work of ADF and MT is supported in part by
the National Science Foundation under grants PHY-0094122 and PHY-0354990, and
by a Cottrell Scholar Award to MT from Research Corporation.


\end{document}